\documentclass[12pt]{article}

\usepackage{latexsym}

\usepackage{amssymb,amsfonts,amsmath}
\usepackage{indentfirst}

 \usepackage{bbm}

\parskip=\medskipamount

\newcommand {\cN}{{\cal N}}

\newcommand {\cV}{{\cal V}}

\def\a{\alpha}

\def\b{\beta}

\def\d{\delta}

\def\g{\gamma}
\def\G{\Gamma}

\def\q{\theta}

\def\x{\xi}
\def\z{\zeta}

\def\F{\Phi}
\def\J{\Psi}
\def\L{\Lambda}

\def\U{\Upsilon}

\def\rd{{\rm d}}
\def\ri{{\rm i}}

\newcommand{\ad}{{\dot{\alpha}}}                           
\newcommand{\bd}{{\dot{\beta}}}                            
\newcommand{\DB}{\bar{D}}

\newcommand{\pa}{\partial}                           
\newcommand{\hf}{\frac12}

\newcommand{\be}{\begin{equation}}
\newcommand{\ee}{\end{equation}}
\newcommand{\bea}{\begin{eqnarray}}
\newcommand{\eea}{\end{eqnarray}}
\newcommand{\non}{\nonumber}
\def\double #1{#1{\hbox{\kern-2pt $#1$}}}

\newcommand{\gd}{{\dot\g}}

\newif\ifdtup

\newcommand{\bsubeq}{\begin{subequations}}
\newcommand{\esubeq}{\end{subequations}}

\numberwithin{equation}{section}

\begin{document}
\begin{titlepage}
\begin{flushright}
October, 2017\\
\end{flushright}
\vspace{5mm}

\begin{center}
{\Large \bf Non-conformal higher spin supercurrents
}\\ 
\end{center}

\begin{center}

{\bf 
Jessica Hutomo and 
Sergei M. Kuzenko
} \\
\vspace{5mm}

\footnotesize{
{\it School of Physics and Astrophysics M013, The University of Western Australia\\
35 Stirling Highway, Crawley W.A. 6009, Australia}}  
~\\
Email: \texttt{20877155@student.uwa.edu.au,
sergei.kuzenko@uwa.edu.au }\\
\vspace{2mm}

\end{center}

\begin{abstract}
\baselineskip=14pt
In four spacetime dimensions
there exist two off-shell  formulations for the massless 
multiplet of superspin $(s+\frac 12)$, where $s=2,3, \dots$.
These supersymmetric higher spin gauge theories, 
known as longitudinal and transverse, are dual to each other 
and  describe two massless fields of spin $(s+\frac 12)$ and $(s+1)$ upon 
elimination of the auxiliary fields. 
They respectively reduce, in the limiting case of $s=1$,
to the linearised actions for the old minimal and the $n=-1$ 
non-minimal ${\cal N}=1$ supergravity theories. 
Associated with these gauge massless theories 
are non-conformal higher spin supercurrent multiplets which we describe.
We demonstrate that the longitudinal higher spin supercurrents 
are realised in the model for a massive chiral scalar superfield
only if $s$ is odd,  $s=2n+1$, with $n= 1,2, \dots$.
\end{abstract}

\vfill

\vfill
\end{titlepage}

\newpage
\renewcommand{\thefootnote}{\arabic{footnote}}
\setcounter{footnote}{0}


\allowdisplaybreaks


\section{Introduction}
\setcounter{equation}{0}

Supercurrent \cite{FZ} is one of the fundamental concepts 
in supersymmetric field theory, for it contains the 
energy-momentum tensor and the supersymmetry current(s).
In the case of superconformal field theories, the supercurrent is unique. 
In particular, the $\cN=1$ conformal supercurrent in four dimensions 
is a real vector superfield $J_{\a\ad}$ subject to the conservation equation 
\cite{FZ}
\bea
\bar D^\ad J_{\a\ad }=0 \quad \Longleftrightarrow \quad 
D^\a J_{\a\ad}=0~.
\label{1.1}
\eea
As an example, we consider the superconformal model for a massless chiral scalar $\F$, $\bar D_\ad \F =0$, with action 
\bea
S_{\rm massless} = \int \rd^4x \rd^2 \q  \rd^2 \bar \q \, \bar \F \F~.
\eea
It is characterised by the supercurrent \cite{FZ}
\bea
J_{\a\ad} = D_\a \F \, \bar D_\ad \bar \F +2\ri (\F\, \pa_{\a\ad} \bar \F 
- \pa_{\a\ad} \F \, \bar \F )~.
\label{1.3}
\eea
The conservation equation \eqref{1.1} follows if one makes use of the equations
of motion $D^2 \F =0$ and $\bar D^2 \bar \F =0$.

In the non-superconformal case, however, 
the conservation equation \eqref{1.1} is replaced by a deformed one. 
Such a deformation is triggered by a trace supermultiplet
containing the trace of the energy-momentum tensor  
 and the $\g$-trace of the supersymmetry current(s).
In general there exist several consistent deformations, and therefore 
several inequivalent supercurrents. 
This means that the problem of classifying inequivalent 
supercurrent multiplets needs to be addressed. 
A simple approach to achieve this 
is to make use of the observation that consistent supercurrents 
are automatically associated with 
linearised off-shell supergravity actions.\footnote{This approach is explained
in detail in \cite{K-var,BK_supercurrent}
for the cases of $\cN=1$ and $\cN=2$ supersymmetric theories in four 
dimensions,  
and in \cite{KT-M11} for $\cN=2$ supersymmetric theories in three dimensions.} 
Given a linearised off-shell action for $\cN=1$ supergravity, 
the  supercurrent conservation equation is obtained by coupling  the supergravity 
prepotentials  to external sources and then demanding the resulting action to be invariant under the linearised supergravity gauge transformations.
Since the linearised off-shell $\cN=1$ supergravity actions have been classified \cite{GKP},
all minimal consistent supercurrents are readily derivable \cite{K-var}.
Reducible supercurrents, such as the $S$-multiplet of \cite{KS2}, 
can be obtained by combining some of the minimal ones. 

The Ferrara-Zumino supercurrent \cite{FZ} is described by the conservation equation
\bea
\bar D^\ad J_{\a\ad }= D_\a T ~, \qquad \bar D_\ad T =0~. 
\label{1.4}
\eea
It corresponds to the old minimal formulation \cite{WZ,old1,old2}
for $\cN=1$ supergravity. An example of a supersymmetric theory 
in which this supercurrent is  realised is the massive chiral model
\bea
S_{\rm massive} = \int \rd^4x \rd^2 \q  \rd^2 \bar \q \, \bar \F \F
+\Big\{ \frac{m}{2} \int \rd^4x \rd^2 \q\, \F^2 +{\rm c.c.} \Big\}~.
\label{1.5}
\eea
For this model,  $J_{\a\ad}$  can be chosen to have the same functional form as in the massless case,  eq. \eqref{1.3}.\footnote{This follows from the fact that the gravitational superfield does not couple to the superpotential \cite{Siegel}.}
The trace multiplet is then given by
\bea
T = m \F^2~.
\label{1.6}
\eea

The higher spin extension of the conformal supercurrent \eqref{1.1} 
was given in \cite{HST}. It is 
\bea
 D^\b J_{\b \a_1 \dots \a_{s-1} \ad_1 \dots \ad_s }=0 
 \quad \Longleftrightarrow \quad 
 \bar D^\bd J_{\a_1 \dots \a_s \bd \ad_1 \dots \ad_{s-1}}=0~,
 \label{1.7}
 \eea
 where $J_{\a(s) \ad(s)} = J_{\a_1 \dots \a_s \ad_1 \dots \ad_s} 
= J_{(\a_1 \dots \a_s)( \ad_1 \dots \ad_s )} $
 is a real superfield. 
This conservation equation is superconformal provided 
 the supercurrent $J_{\a(s) \ad(s)}$ is 
superconformal primary of weight $(1+ \frac{s}{2},1+ \frac{s}{2})$
\cite{KMT}.
The higher spin extension of the massless supercurrent \eqref{1.3}
was given in \cite{KMT}. It is 
\bea
J_{\a(s) \ad (s)} &=&   (2\ri)^{s-1} \sum_{k=0}^{s} (-1)^k \binom{s}{k}\non \\
&&\times  \left\{ \binom{s}{k+1} 
 \pa_{(\a_1 ( \ad_1} \dots \pa_{\a_k \ad_k} D_{\a_{k+1}} \F \,
\bar D_{\ad_{k+1} } \pa_{\a_{k+2} \ad_{k+2} }\dots \pa_{ \a_s) \ad_s  ) } \bar \F 
\right.\non \\
&& \qquad \left.+ 2\ri \binom{s}{k} 
\pa_{(\a_1  (\ad_1} \dots \pa_{\a_k \ad_k}  \F \,
 \pa_{\a_{k+1} \ad_{k+1} }\dots \pa_{ \a_s) \ad_s  ) } \bar \F \right\}~.
 \label{1.8}
\eea
This supercurrent was also re-derived in a revised version (v2, 10 Oct) of \cite{BGK}.

To obtain higher spin extensions of non-conformal supercurrents, 
one can make use of the known gauge off-shell formulations for  massless 
higher spin supermultiplets. Such formulations were developed 
in the early 1990s in Minkowski superspace \cite{KSP,KS}
and anti-de Sitter superspace \cite{KS94}
(see \cite{Ideas} for a pedagogical review of the results of  \cite{KSP,KS}).
In section 2 we briefly review the two formulations for each massless 
multiplet of half-integer superspin $s+1/2$, with $s=2, 3, \ldots$. 
In section 3 we present the non-conformal higher spin supercurrent
multiplets associated with these gauge theories. 
As an application, we derive all higher spin supercurrents 
for the massive model \eqref{1.5}.


\section{Massless half-integer superspin multiplets}

For a  massless multiplet of half-integer superspin $s+1/2$, with $s=2, 3, \ldots$,
there exist two off-shell formulations \cite{KSP}
which are referred to as 
transverse and longitudinal. They are 
described in terms of 
the following dynamical variables:
\begin{subequations}
\bea
\cV^\bot_{s+1/2}& = &\Big\{H_{\a(s)\ad(s)}~, ~
\G_{\a(s-1) \ad(s-1)}~,
~ \bar{\G}_{\a(s-1) \ad(s-1)} \Big\} ~,    \\
\label{10}
\cV^{\|}_{s+1/2} &=& 
\Big\{H_{\a(s)\ad(s)}~, ~
G_{\a(s-1) \ad(s-1)}~,
~ \bar{G}_{\a(s-1) \ad(s-1)} \Big\}
~.
\eea
\end{subequations}
Here $H_{\a(s) \ad (s)}$ is a real unconstrained superfield.
The complex superfields 
$\G_{\a (s-1) \ad (s-1)} $ and 
$G_{\a (s-1) \ad (s-1)}$ are  transverse linear and 
longitudinal linear in the sense that they obey the constraints\footnote{More
generally, complex tensor superfields $\G_{\a(r) \ad(t)}$ and $G_{\a(r) \ad(t)}$ are called  transverse linear and longitudinal linear, respectively,
 if the constraints
$\bar D^\bd \G_{ \a(r) \bd \ad(t - 1) } = 0 $ and
$\bar D_{(\bd} G_{\a(r)\ad(t) )} = 0 $ are satisfied. 
The former constraint is defined for $t \neq 0$; it has to be replaced 
with the standard linear 
constraint,  $\bar D^2 \G_{ \a(r) } = 0 $, for $t=0$.
The latter constraint for $t = 0$ is the chirality condition
$\bar D_\bd G_{ \a(r) } = 0$.
}
\begin{subequations}
\bea
{\bar D}^\bd \,\G_{\a(s-1) \bd \ad(s-2)} =  0 \quad
&\Longrightarrow & \quad \bar D^2 \G_{\a(s-1) \ad(s-1)}=0~,\label{transverse}
\\
{\bar D}_{ (\ad_1} \,G_{\a(s-1) \ad_2 \dots \ad_{s-1})}=0  \quad
&\Longrightarrow & \quad \bar D^2 G_{\a(s-1) \ad(s-1)}=0~,
\label{longitudinal}
\eea
\end{subequations}
These constraints can be solved in terms of unconstrained prepotentials
as follows:
\begin{subequations}
\bea
 \G_{\a(s-1) \ad(s-1)}&=& \bar D^\bd 
{ \Phi}_{\a(s-1)\,(\bd \ad_1 \cdots \ad_{s-1}) } ~,
\label{2.2a}
 \\
 G_{\a(s-1) \ad(s-1)} &=& {\bar D}_{( \ad_1 }
 \Psi_{ \a(s-1) \, \ad_2 \cdots \ad_{s-1}) } ~.
\label{2.2b}
\eea
\end{subequations}
The prepotentials  are defined modulo gauge transformations of the form:
\begin{subequations}
\bea
\d_\x \Phi_{\a(s-1)\, \ad (s)} 
&=&  \bar D^\bd 
{ \x}_{\a(s-1)\, (\bd \ad_1 \cdots \ad_{s}) } ~,
\label{tr-prep-gauge}
\\
\d_\z  \Psi_{ \a(s-1) \, \ad {(s-2}) } &=&  {\bar D}_{( \ad_1 }
 \z_{ \a(s-1) \, \ad_2 \cdots \ad_{s-2}) } ~,
\label{lon-prep-gauge}
\eea
\end{subequations}
with the gauge parameters $ {\x}_{\a(s-1)\,  \ad (s+1) } $
and $ \z_{ \a(s-1) \, \ad (s-3)}$ being unconstrained.\footnote{For $s=2$
the gauge transformation law \eqref{lon-prep-gauge} has to be replaced with 
 $\d \J = \z$, with the gauge parameter $\z$ being chiral, $\bar D_\ad \z=0$.}

The gauge transformations of the superfields $H$, $\G$ and $G$ are 
\begin{subequations} \label{2.5}
\bea \d_\L H_{\a_1 \dots \a_s \ad_1  \dots \ad_s} 
&= &\bar D_{(\ad_1} \L_{\a_1\dots  \a_s \ad_2 \dots \ad_s )} 
- D_{(\a_1} \bar{\L}_{\a_2 \dots \a_s)\ad_1  \dots \ad_s} \ , \label{2.5a} \\
\d_\L \G_{\a_1 \dots \a_{s-1}\ad_1 \dots \ad_{s-1}} &= &
-\frac{1}{4} \bar D^\bd D^2 \bar{\L}_{\a_1 \dots \a_{s-1}\bd \ad_1 
\dots \ad_{s-1}} \ , \label{2.5b} \\
\d_\L G_{\a_1 \dots \a_{s-1}\ad_1 \dots \ad_{s-1}} &= & - \hf \bar D_{(\ad_1} 
\bar D^{|\bd|} D^\b \L_{\b\a_1 \dots \a_{s-1} \ad_2 \dots \ad_{s-1}) \bd} \non\\
&&+ \ri (s-1) \bar D_{(\ad_1} \pa^{\b |\bd|} 
\L_{\b \a_1 \dots \a_{s-1} \ad_2 \dots \ad_{s-1} ) \bd} \ . \label{2.5c}
\eea
\end{subequations}
Here the gauge parameter $\L_{\a_1 \dots \a_s \ad_1 \dots \ad_{s-1}}
=\L_{(\a_1 \dots \a_s )(\ad_1 \dots \ad_{s-1})}$ 
is unconstrained. 
The symmetrisation in \eqref{2.5c} is extended only to the indices 
$\ad_1, \ad_2, \dots,  \ad_{s-1}$. 
It follows from  \eqref{2.5b} and \eqref{2.5c} that the transformation laws of the prepotentials $\F_{\a(s-1) \ad(s)}$ and $\J_{\a(s-1) \ad(s-2)}$ are
\begin{subequations}
\bea
\d_\L \F_{\a_1 \dots \a_{s-1}\ad_1 \dots \ad_s} &= &
-\frac{1}{4} D^2 \bar{\L}_{\a_1 \dots \a_{s-1} \ad_1 
\dots \ad_s} \ , \label{2.6a} \\
\d_\L \J_{\a_1 \dots \a_{s-1}\ad_1 \dots \ad_{s-2}} &= & - \hf 
\Big( \bar D^{\bd} D^\b -2\ri (s-1) \pa^{\b \bd} \Big)
\L_{\b\a_1 \dots \a_{s-1} \bd \ad_1 \dots \ad_{s-2}} ~.
\label{2.6b}
\eea
\end{subequations}

In the transverse formulation, 
the action  invariant 
under the gauge transformations (\ref{2.5a}) and
(\ref{2.5b})  is 
\bea
S^{\bot}_{s+1/2}[H, \G ,\bar \G]&=&
\Big( - \frac{1}{2}\Big)^s  \int \rd^4x \rd^2 \q  \rd^2 \bar \q \,
\Big\{ \frac{1}{8} H^{ \a(s) \ad(s) }  D^\b {\bar D}^2 D_\b 
H_{\a(s) \ad(s) }  \non \\
&+& H^{ \a(s) \ad(s) }
\left( D_{\a_s}  {\bar D}_{\ad_s} \G_{\a(s-1) \ad(s-1) }
- {\bar D}_{\ad_s}  D_{\a_s} 
{\bar \G}_{\a (s-1) \ad (s-1) } \right) \non \\
&+&\Big( {\bar \G} \cdot \G
+ \frac{s+1} {s} \, \G \cdot \G ~+~ {\rm c.c.} \Big)
\Big\} ~.
\label{hi-t}
\eea
In the longitudinal formulation, the action  invariant 
under the gauge transformations (\ref{2.5a}) and
(\ref{2.5c}) 
is
\bea
S^{\|}_{s+1/2}[H,G,\bar G]&=&
\Big(-\hf \Big)^s  \int \rd^4x \rd^2 \q  \rd^2 \bar \q \, \Big\{
\frac 18 
H^{\a (s) \ad (s) }   D^\b \bar D^2
 D_\b H_{\a (s) \ad (s) }  \non \\
&-& \frac{1}{8} \, \frac{s}{2s+1} \, \Big( \,
\big[ D_{\g}, \bar D_{\gd}\big] H^{\g \a (s-1)\gd  \ad (s-1)}
\, \Big)  \,
\big[ D^\b, \bar D^{\bd}\big]
H_{\b\a(s-1)\bd\ad(s-1)} \,  \non \\
&+& \frac{s}{2}\, \Big( \partial_{\g\gd} 
H^{\g \a (s-1) \gd \ad (s-1)}  \Big) \,
\partial^{\b\bd}
H_{\b\a(s-1)\bd\ad(s-1)} 
\non \\
&+& 2{\rm i} \, \frac{ s}{2s+1}  \,  \pa_{\g \gd } 
H^{\g \a (s-1) \gd \ad (s-1)}
\Big( G_{\a(s-1) \ad(s-1)} - \bar G_{\a(s-1) \ad(s-1)} \Big)  \non \\
&+& \frac{1}{2s+1} \Big( \bar G \cdot G - \frac{s+1}s G \cdot G
+ {\rm c.c.}\Big)\Big\}   ~.
\label{hi-l}
\eea
The models (\ref{hi-t}) and (\ref{hi-l}) are dually equivalent 
\cite{KSP}.

We now briefly comment on the limiting $s=1$ case which should correspond to supergravity. The transverse linear constraint \eqref{transverse} cannot be used
for $s=1$, however its corollary $\bar D^2 \G_{\a(s-1) \ad(s-1)}=0$ can be used,
\bea
\bar D^2 \G =0~.
\eea 
This constraint defines a complex linear superfield. In accordance with \eqref{2.5b}, 
the gauge transformation of $\G$ is 
\bea
\d_\L \G = \frac{1}{4} \bar D_\bd D^2 \bar{\L}^\bd ~.
\eea
The action \eqref{hi-t} for $s=1$ coincides with the linearised action 
for the $n=-1$ non-minimal supergravity, see \cite{GKP,Ideas} for reviews.

The longitudinal linear constraint  
\eqref{longitudinal} 
is the chirality condition for $s=1$, 
\bea
\bar D_\ad G=0~.
\eea
The gauge transformation law \eqref{2.5c} cannot directly be used for $s=1$. 
Nevertheless,  it can be rewritten in the form 
\bea
\d_\L G_{\a_1 \dots \a_{s-1}\ad_1 \dots \ad_{s-1}} &= &
 - \frac 14  \bar D^2  
 D^\b \L_{\b\a_1 \dots \a_{s-1} \ad_1 \dots \ad_{s-1}} \non\\
&&+ \ri (s-1)  \pa^{\b \bd} \bar D_{(\ad_1}
\L_{\b \a_1 \dots \a_{s-1} \ad_2 \dots \ad_{s-1} ) \bd} ~, 
\eea
which is well defined for  $s=1$:
\bea
\d_\L G &= & - \frac 14  \bar D^2   D^\b \L_\b~.
\eea
The action \eqref{hi-l} for $s=1$ coincides with the linearised action 
for the old minimal supergravity, see \cite{GKP,Ideas} for reviews.
There are actually three different realisations for $G$ in terms of unconstrained 
superfields (see also \cite{FLMS,KT-M17} for recent discussions). 
The standard realisation is 
\bea
G = -\frac{1}{4} \bar D^2 U~, 
\eea
where the prepotential $U$ is an unconstrained 
complex superfield, with the gauge transformation law 
$\d_\L U = D^\b \L_\b$. It is this realisation which corresponds to the old minimal supergravity. The second realisation is to make use of a three-form multiplet \cite{Gates}
\bea
G=-\frac{1}{4}\DB^2P~,~~~~~~
\bar{P}= P
~,
\eea
where $P$ is a real but otherwise unconstrained prepotential, with the gauge 
transformation law $\d_\L P =  D^\b \L_\b +\bar D_\bd \bar \L^\bd$.
This realisation corresponds to the so-called three-form supergravity \cite{GS}.
Finally,  $G$ can be chosen to be  a complex three-form multiplet \cite{GS}
\bea
G=-\frac{1}{4}\bar D^2 D^\b \U_\b~,~~~~~~
\eea
where $\U_\b $ is an unconstrained  complex spinor prepotential, 
with the gauge transformation $\d_\L \U_\b = \L_\b$.
This realisation corresponds to the  complex 
three-form supergravity \cite{old1}.


\section{Higher spin supercurrents}

In this section we first describe the general structure of non-conformal higher spin supercurrents \cite{KS-unpublished}.

In the framework of the longitudinal formulation, 
let us couple the prepotentials 
$H_{ \a (s) \ad (s) } $, $\J_{ \a (s-1) \ad (s-2) }$ and $\bar \J_{ \a (s-2) \ad (s-1) }  $,
to external sources
\bea
S^{(s+\hf)}_{\rm source}=\int \rd^4x \rd^2 \q  \rd^2 \bar \q \, \Big\{ 
H^{ \a (s) \ad (s) } J_{ \a (s) \ad (s) }
&+& \J^{ \a (s-1) \ad (s-2) } T_{ \a (s-1) \ad (s-2) } \non \\
&+& \bar \J_{ \a (s-2) \ad (s-1) } \bar T^{ \a (s-2) \ad (s-1) } \Big\}~.
\label{3.1}
\eea
Requiring $S^{(s+\hf)}_{\rm source}$ to be invariant under 
\eqref{lon-prep-gauge} gives
\begin{subequations} \label{3.3}
\bea
\bar D^\bd T_{\a(s-1) \bd \ad_1 \dots \ad_{s-3}} =0~,
\label{3.2}
\eea
and therefore $T_{ \a (s-1) \ad (s-2) } $ is a transverse linear superfield. 
Requiring $S^{(s+\hf)}_{\rm source}$ to be invariant under the gauge transformations
(\ref{2.5a}) and (\ref{2.6b}) gives the following conservation equation:
\bea
\bar D^\bd J_{\a_1 \dots \a_s \bd \ad_1 \dots \ad_{s-1}} 
+\hf \Big( D_{(\a_1} \bar D_{(\ad_1}
-2\ri (s-1) \pa_{ (\a_1 (\ad_1 } \Big)  T_{\a_2\dots \a_s) \ad_2 \dots \ad_{s-1})} =0~.
\label{3.3a}
\eea
For completeness, we also give the conjugate equation
\bea
D^\b J_{\b \a_1 \dots \a_{s-1}  \ad_1 \dots \ad_{s}} 
-\hf \Big( \bar D_{(\ad_1}  D_{(\a_1}
-2\ri (s-1) \pa_{ (\a_1 (\ad_1 } \Big)  \bar T_{\a_2\dots \a_{s-1}) \ad_2 \dots \ad_{s})} 
=0~. \label{3.3c}
\eea
\end{subequations}

Similar considerations for the transverse formulation lead to the following non-conformal supercurrent multiplet
\begin{subequations}\label{TSupercurrent}
\bea
\bar D^\bd J_{\a_1 \dots \a_s \bd \ad_1 \dots \ad_{s-1}} 
-\frac{1}{4} \bar D^2 F_{\a_1 \dots \a_s \ad_1 \dots \ad_{s-1}} &=&0~,\\
D_{(\a_1 } F_{\a_2 \dots \a_{s+1} )\ad_1 \dots \ad_{s-1}}&=&0  ~.
\eea
\end{subequations}
Thus  the trace multiplet $\bar F_{\a(s-1) \ad(s)}$ is longitudinal linear. 

In the remainder of this section we are going to show that it is the longitudinal 
higher spin supercurrents \eqref{3.3} which naturally arise
in the massive chiral model \eqref{1.5}. 
As in \cite{GK}, it is useful to introduce 
auxiliary complex variables $\z^\a \in {\mathbb C}^2$ and their conjugates 
$\bar \z^\ad$. Given a tensor superfield $U_{\a(p) \ad(q)}$, we associate with it 
the following  field on ${\mathbb C}^2$ 
\bea
U_{(p,q)} (\z, \bar \z):= \z^{\a_1} \dots \z^{\a_p} \bar \z^{\ad_1} \dots \bar \z^{\ad_q}
U_{\a_1 \dots \a_p \ad_1 \dots \ad_q}~,
\eea
which is homogeneous of degree $(p,q)$ in the variables $\z^\a$ and $\bar \z^\ad$.
We introduce operators that  increase the degree 
of homogeneity in the variables $\z^\a$ and $\bar \z^\ad$, 
\bea
{D}_{(1,0)} := \z^\a D_\a~, \qquad 
{\bar D}_{(0,1)} := \bar \z^\ad \bar D_\ad~, 
\qquad 
{\pa}_{(1,1)} := 2\ri \z^\a \bar \z^\ad \pa_{\a\ad}~,
\eea
and their descendants 
\bea
A_{(1,1)} := -D_{(1,0)} \bar D_{(0,1)} +(s-1) \pa_{(1,1)} ~, \quad 
\bar A_{(1,1)} := \bar D_{(0,1)}  D_{(1,0)} -(s-1) \pa_{(1,1)} ~. 
\eea
The fermionic operators ${D}_{(1,0)} $ and 
${\bar D}_{(0,1)} $ are nilpotent, ${D}_{(1,0)}^2=0 $ and 
${\bar D}_{(0,1)}^2=0 $.
We also introduce two {\it nilpotent} operators that decrease the degree 
of homogeneity in the variables $\z^\a$ and $\bar \z^\ad$, specifically
\begin{subequations}
\bea
D_{(-1,0)} &:=& D^\a \frac{\pa}{\pa \z^\a}~, \qquad D_{(-1,0)}^2 =0~,\\
\bar D_{(0,-1)}& :=& \bar D^\ad \frac{\pa}{\pa \bar \z^\ad}~ \qquad 
\bar D_{(0,-1)}^2 =0 
~.
\eea
\end{subequations}

Making use of the notation introduced, 
the transverse linear condition \eqref{3.2} and its conjugate become
\begin{subequations}
\bea
\bar D_{(0,-1)} T_{(s-1,s-2)} &=&0~,  \label{3.8a}\\
D_{(-1,0)} \bar T_{(s-2,s-1)} &=&0~.
\eea
\end{subequations}
The conservation equations \eqref{3.3a} and \eqref{3.3c} turn into 
\begin{subequations}
\bea
\frac{1}{s}\bar D_{(0,-1)} J_{(s,s)} -\hf A_{(1,1)} T_{(s-1, s-2)}&=&0~, \label{3.9a}\\
\frac{1}{s} D_{(-1,0)} J_{(s,s)} -\hf \bar A_{(1,1)} \bar T_{(s-2, s-1)}&=&0~.
\eea
\end{subequations}
Since the operator $\bar D_{(0,-1)} J_{(s,s)} $ is nilpotent, 
the conservation equation \eqref{3.9a} is consistent provided
\bea
\bar D_{(0,-1)}  A_{(1,1)} T_{(s-1, s-2)}=0~.
\eea
This is indeed true, as a consequence of the transverse linear condition
\eqref{3.8a}.

Using the notation introduced, the massless higher spin supercurrent \eqref{1.8}
becomes
\bea
J_{(s,s)} &=& \sum_{k=0}^s (-1)^k
\binom{s}{k}
\left\{ \binom{s}{k+1} 
{\pa}^k_{(1,1)}
 D_{(1,0)} \F \,
{\pa}^{s-k-1}_{(1,1)}
\bar D_{(0,1)} 
\bar \F  
\right. \non \\ 
&& \left.
 \qquad \qquad
+ \binom{s}{k} 
{\pa}^k_{(1,1)}
  \F \,
{\pa}^{s-k}_{(1,1)}
\bar \F \right\}
\label{3.12}
\eea

We now turn to constructing non-conformal higher spin supercurrents
arising in the massive model \eqref{1.5}. Guided by the structure of  
the Ferrara-Zumino supercurrent for the model \eqref{1.5},
we assume that $J_{(s,s)} $ has the same functional form as in the massless
case, eq. \eqref{3.12}. Making use of the massive equation of motion, 
\bea
-\frac{1}{4} \bar D^2 \bar \F +m \F =0~,
\eea 
we obtain
\begin{subequations}
\bea
\bar D_{(0,-1)} J_{(s,s)} &=& F_{(s,s-1)}~, \label{3.12a}
\eea
where we have denoted 
\bea
F_{(s,s-1)} &=& 2m(s+1) \sum_{k=0}^s (-1)^{s-1+k} \binom{s}{k} \binom{s}{k+1}
\non \\ 
&& 
\times \left\{1+(-1)^s \frac{k+1}{s-k+1}\right\} 
 {\pa}^k_{(1,1)} \F \,{\pa}^{s-k-1}_{(1,1)}
 D_{(1,0)} \F ~.
\eea
\end{subequations}

Keeping in mind eq. \eqref{3.9a}, 
 we now  look for a superfield $T_{(s-1, s-2)}$
such that (i) it obeys the transverse linear constraint \eqref{3.8a}; and 
(ii) it satisfies the equation
\bea
F_{(s,s-1)} = \frac{s}{2} A_{(1,1)} T_{(s-1, s-2)}~. 
\eea
We consider a general ansatz
\bea
T_{(s-1, s-2)} = (-1)^s m \sum_{k=0}^{s-2} c_k 
{\pa}^k_{(1,1)} \F\,
{\pa}^{s-k-2}_{(1,1)}
 D_{(1,0)} \F ~.
 \label{T3.15}
\eea
For $k = 1,2,...s-2$, condition (i) implies that 
the coefficients $c_k$ must satisfy
\begin{subequations}\label{3.16}
\begin{align}
kc_k = (s-k-1) c_{s-k-1}~,\label{3.16a}
\end{align}
while (ii) gives the following equation
\begin{align}
c_{s-k-1} + s c_k + (s-1) c_{k-1} &= -4(-1)^k \frac{s+1}{s} \binom{s}{k} \binom{s}{k+1} 
\non \\
& \qquad \qquad \times \left\{ 1+ (-1)^s \frac{k+1}{s-k+1} \right\} ~.\label{3.16b}
\end{align}
Condition (ii) also implies that 
\begin{align}
(s-1) c_{s-2} +c_0 &= 4(-1)^s (s+1)\left\{1+(-1)^s \frac{s}{2}\right\}~, \label{3.16c}\\
c_0 &= -\frac{4}{s}(s+1+(-1)^s)~. \label{3.16d}
\end{align}
\end{subequations}
It turns out that the equations \eqref{3.16} 
lead to a unique expression for $c_k$ given by 
\bea\label{3.17}
c_k &=& -\frac{4(s+1)(s-k-1)}{s(s-1)}
\sum_{l=0}^k \frac{(-1)^k}{s-l} \binom{s}{l} \binom {s}{l+1} \left\{ 1+(-1)^s \frac{l+1}{s-l+1} \right\}  ~,~~~~  \\
&& \qquad \qquad \qquad  k=1,2,\dots s-2~. \non 
\eea
If the parameter $s$ is odd, $s=2n+1$, with  $n=1,2,\dots$, 
one can check that the equations \eqref{3.16a}--\eqref{3.16c} are identically 
satisfied. 
However, if the parameter $s$ is even, $s=2n$, with $n=1,2,\dots$, 
there appears an inconsistency: 
 the right-hand side of \eqref{3.16c} is positive, while the left-hand side 
is negative, $(s-1) c_{s-2} + c_0 < 0$. Therefore, our solution \eqref{3.17} is only consistent for $s=2n+1, n=1,2,\dots$. 

Relations \eqref{3.12}, \eqref{T3.15}, \eqref{3.16d} and \eqref{3.17} determine the non-conformal higher spin supercurrents 
in the massive chiral model \eqref{1.5}, with the trace multiplet
$T_{(s-1, s-2)}$
being the higher spin extension of \eqref{1.6}.
Unlike the conformal higher spin supercurrents \eqref{1.8},
the non-conformal ones exist only for the odd values of $s$,
$s=2n+1$, with  $n=1,2,\dots$.

\section{Concluding comments}

The non-conformal higher spin supercurrent multiplets 
\eqref{3.3} and \eqref{TSupercurrent}
are automatically consistent, since they are associated with the gauge-invariant 
models (\ref{hi-l}) and (\ref{hi-t}), respectively. 
An interesting open question is to classify all non-conformal deformations 
of the higher spin supercurrents \eqref{1.7}, along the lines of the recent 
analysis of non-conformal $\cN=(1,0)$ supercurrents in six dimensions
\cite{KNT}.
Our results provide the setup required for developing a program to derive higher spin supersymmetric models from quantum correlation functions,  as an extension of the non-supersymmetric approaches pursued, e.g., in \cite{Bonora1,Bonora2,Bonora3}.
Our results also have a natural extension to the case of $\cN=2$ supersymmetry
in three dimensions, where the off-shell higher spin supermultiplets
have recently been constructed in \cite{KO}.

Shortly before posting this work to the arXiv, there appeared another revised version (v3, 26 Oct)
of Ref. \cite{BGK} containing a new section devoted to the higher spin supercurrents 
in the massive chiral model \eqref{1.5}.
These authors also observed that the  higher spin supercurrents $J_{\a(s) \ad(s)}$
in the massive chiral model \eqref{1.5} exist only for the  odd values of $s$, 
$s=2n+1$, with  $n=1,2,\dots$


\noindent
{\bf Acknowledgements:}\\
SMK is grateful to Darren Grasso for comments on the manuscript. 
The work of JH is supported by an Australian Government Research Training Program (RTP) Scholarship.
The work of SMK is supported in part by the Australian 
Research Council, project No. DP160103633.


\begin{footnotesize}

\end{footnotesize}

\end{document}